# Urine dataset having eight particles classes

Taner Tuncer[1,*], Merve Erkuş[1], Ahmet Çınar[1], Hakan Ayyıldız[2], Seda Arslan Tuncer[3]

[1]Fırat University, Faculty of Engineering, Computer Engineering, 23119, Elazığ, Turkey.
[2]Elazığ Fethi Sekin Central Hospital, 23000, Elazığ, Turkey
[3]Fırat University, Faculty of Engineering, Software Engineering, 23119, Elazığ, Turkey.
[1,*]ttuncer@firat.edu.tr

*Abstract*— Urine sediment examination (USE) is one of the main tests used in the evaluation of diseases such as kidney, urinary, metabolic, and diabetes and to determine the density and number of various cells in the urine. USE's manual microscopy is a labor-intensive and time-consuming, imprecise, subjective process. Recently, automatic analysis of urine sediment has become inevitable in the medical field. In this study, we propose a dataset that can be used by artificial intelligence techniques to automatically identify particles in urine sediment images. The data set consists of 8509 particle images obtained by examining the particles in the urine sediment obtained from 409 patients from the Biochemistry Clinics of Elazig Fethi Sekin Central Hospital. Particle images are collected in 8 classes in total and these are Erythrocyte, Leukocyte, Epithelial, Bacteria, Yeast, Cylinders, Crystals, and others (sperm, etc.).

## I. Specification Table

Subject: Biomedical Image Processing
Specific subject area: A compiled dataset of eight different particle images from urine sediment images.
Type of data: Image
How the data were acquired: Images were obtained from a total of 409 patients who applied to Elazig Fethi Sekin Central Hospital. The precipitated urine sediment from each patient was compiled from the microscope image by labeling the most common particles. Images were obtained from the Optika B293PLi microscope.
Data format: jpg
Data accessibility: Public

## II. Value of Data

This article presents a dataset and its properties that include images of urine sediment and particles in the image collected manually via a binocular microscope. The main features of the created dataset are as follows.

1. This dataset is useful for identifying particles in urine images and for training machine learning models [2].

2. Useful for classifying possible particles in urine images through Deep Learning (DL) techniques.

3. There are particle images of 8 different classes in the data set.

4. The dataset can be used as educational dataset for biochemistry laboratory workers as well as a unique resource for a better understanding of particles.

## III. Objective

Kidneys, one of the organs of the excretory system, are the organs that regulate the amount of water in our body by filtering unwanted substances in the blood, protecting the electrolytes and proteins that the body will use again, and thus ensuring the balance of the body [1]. Substances that are not deemed necessary by the body are excreted from the kidneys and transmitted to the bladder and are excreted out of the body by passing through the urethra. This liquid, which is thrown out of the body and called urine, is used in the diagnosing and follow-up of many diseases such as blood, kidney, urinary, metabolic, and diabetes. For this aim, examinations are made regarding the presence of substances that should not be in the urine. This examination, which is widely used and frequently requested by doctors, is known as the Complete Urine Test. In this test, many parameters such as density, cylinder, Ph, Bilirubin, erythrocyte, glucose, color, clarity, leukocyte esterase, ketone, nitrite, protein, urobilinogen, and bacteria are examined. One or more of these parameters indicate that the individual is healthy or unhealthy according to the value they show. The traditional method for microscopic examination of urine sediment is manual examination. The fact that the particles in the urine images contain many differences and the images are complex makes this process difficult.

## IV. Experimental Setup and Data Description

With the approval of the ethics committee with the document dated 20.05.2022 and numbered 8455, a total of 409 urine samples, 143 women and 266 men, who applied to Elazig Fethi Sekin Research and Training Hospital, were collected. The urine sample was taken for examination in the first 1 hour. Microscopic examination was carried out in accordance with the procedure outlined below.

Step 1: 10 ml of urine sample is put into a centrifuge tube.

Step 2: It is centrifuged at 1500 rpm for 5 minutes and the urine sediment is precipitated.

Step 3: The remaining liquid part is poured out.

Step 4. The underlying sediment is taken on a clean slide and examined under a microscope after covering the coverslip.

Figure 1.a and Figure 1.b show the experimental setup for obtaining urine image and a urine image taken from the microscope, respectively.

Images were taken with the Optika B293PLi microscope with Trinocular brightfield and a camera device with High speed, user-friendly camera with high resolution (10 MP), CMOS sensor, and USB3.0 connection. Each analyzed image is in 3664x2748 color JPG format. The total number of particles examined on the images obtained is 8509, and the number of particles according to their classes is as in Table 1. Labeling of each particle was carried out by a specialist in biochemistry.

TABLE I.  PARTICLE CLASSES AND NUMBERS

| Particle | Number of particles |
|---|---|
| Erythrocyte | 2279 |
| Leukocyte | 1734 |
| Epithelium | 432 |
| Bacteria | 1224 |
| Yeast | 688 |
| Cylinders | 240 |
| Crystals | 1842 |
| Other | 70 |
| Total | 8509 |

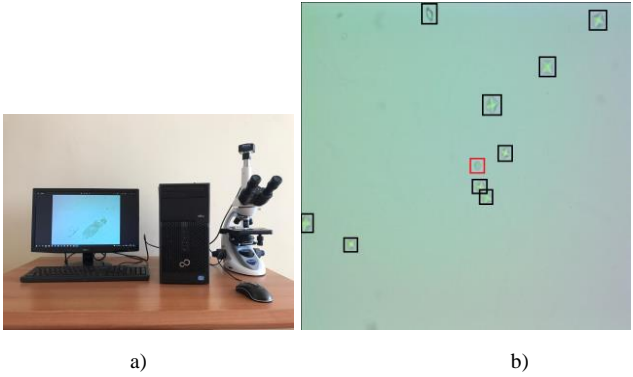

Fig. 1. a) Experimental setup, b) a urine image

The particles in Table 1 can be detailed as follows.

Erythrocyte (RBC Red Blood Cell): Normally, there are a few RBCs (Normally 0-5 RBCs) in the urine sediment. RBC found on microscopic examination indicates blood in the urine. The presence of blood in the urine is not a normal situation, and the underlying cause of this situation should be investigated and blood tests should be performed. If there is blood in the urine along with bacteria and leukocytes, it may be a urinary tract infection that can be easily treated with antibiotics. Figure 2 shows Erythrocyte, Leukocyte, Epithelial, Bacteria, Yeast, Cylinders, Crystals, and others particles in the urine sample, respectively.

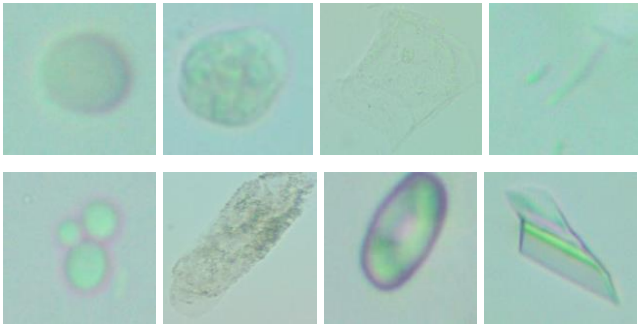

Fig. 2. Particles in the urine

Leukocyte (WBC; White Blood Cell): The WBC count in the urine sediment is low in normal individuals and is between 0 and 5. An increased WBC count in the urine examined under the microscope indicates inflammation or infection anywhere in the urinary tract. If it is seen together with bacteria, it may be the result of a possible urinary tract infection.

Epithelial: In the urine analysis, these cells can be seen as few, medium or many. It is normal to have a few epithelial cells in the urinary sediment of women and men. An increased number of epithelial cells is present in conditions such as infections, inflammation, and malignancies.

Bacteria: The urinary tract of healthy people is sterile. Normally, bacteria are not found in the urine sediment under the microscope. If there are bacteria in the urine, it will be positive and there is a pathological condition. As a result, the presence of bacteria is a harbinger of urinary tract infections. A urine culture is requested for further examination. If bacteria are seen in the urine sediment, it is recorded in the complete urinalysis report as little, medium, or high. If the infection is not treated, it can be carried to the kidneys and cause kidney disease. Although urinary tract infections can be easily treated, if the person is constantly recurring, a urine culture and sensitivity test are requested to help with the treatment.

Yeast: Rarely in men, more often in women, yeast can be found in the urine. It is more common in women with vaginal yeast infections. Because at the time of collection, the urine is contaminated with vaginal secretions. If yeast is observed in the urine, the person may be treated for yeast infection.

Cylinder: These cylindrical particles consist of coagulated protein released into the urine by kidney cells. It occurs in the long, thin, hollow ducts of the kidneys. They usually take the shape of a tubule. When viewed under a microscope it often resembles a sausage shape and is almost clear in healthy people. Such casts are called hyaline casts. Normally healthy people may have several hyaline casts per low power field. More hyaline casts may occur if strenuous exercise is performed. Other types of rollers are associated with different kidney diseases. The caster types in the urine can give an idea about which disorder affects the kidneys.

Crystal: Urine has many dissolved substances. These are waste chemicals that the body needs to eliminate. These solutes can form crystals of a particular substance in the urine in the following cases:

- The concentration of dissolved substances is increased.
- Urine pH becomes more and more acidic or basic.
- Urine heat promotes their formation.

Crystals are identified by their color, shape, and urine pH. They may be without definite shapes, such as small, sand-like particles, or they may have special needle-like shapes. They are considered normal if they consist of solutes typically found in the urine.

## V. CONCLUSION

We hope that the created dataset will provide a good reference for automated use as well as cell detection and can be used for training to take a big step forward in practice. In addition, the urine data set that we have made public in the study will expand the scope of existing data sets and contribute to new research in this direction as it includes more classes.

**Conflict of interest statement**: The authors declare no competing interests.

**Ethical statement**: The data set used in this study was collected with the decision of the ethics committee dated 20.05.2022 and numbered 8455.

**Funding:** This work was supported Scientific And Technological Research Institution Of Turkey. Project No. 122E094.